\documentclass[fleqn,usenatbib]{mnras}

\usepackage{newtxtext,newtxmath}

\usepackage[T1]{fontenc}

\DeclareRobustCommand{\VAN}[3]{#2}
\let\VANthebibliography\thebibliography
\def\thebibliography{\DeclareRobustCommand{\VAN}[3]{##3}\VANthebibliography}

\usepackage{graphicx}	
\usepackage{amsmath}	
\usepackage{amssymb}	

\newcommand{\ie}{\textit{i}.\textit{e}.}

\newcommand{\new}{\color{black}}

\title[Black Dwarf Supernova in the Far Future]{Black Dwarf Supernova in the Far Future}

\author[M. E. Caplan et al.]{
M. E. Caplan,$^{1}$\thanks{E-mail: mecapl1@ilstu.edu}
\\
$^{1}$Department of Physics, Illinois State University, Normal, IL 61761 USA\\
}

\date{Accepted XXX. Received YYY; in original form ZZZ}

\pubyear{2020}

\begin{document}
\label{firstpage}
\pagerange{\pageref{firstpage}--\pageref{lastpage}}
\maketitle

\begin{abstract}

In the far future long after star formation has ceased the universe will be populated by sparse degenerate remnants, mostly white dwarfs, though their ultimate fate is an open question. These white dwarfs will cool and freeze solid into black dwarfs while pycnonuclear fusion will slowly process their composition to iron-56. However, due to the declining electron fraction the Chandrasekhar limit of these stars will be decreasing and will eventually be below that of the most massive black dwarfs. As such, isolated dwarf stars with masses greater than $\sim 1.2 M_\odot$ will collapse in the far future due to the slow accumulation of iron-56 in their cores. If proton decay does not occur then this is the ultimate fate of about $10^{21}$ stars, approximately one percent of all stars in the observable universe. We present calculations of the internal structure of black dwarfs with iron cores as a model for progenitors. 
From pycnonuclear fusion rates we estimate their lifetime and thus delay time to be $10^{1100}$ years. We speculate that \new{high mass black dwarf supernovae resemble accretion induced collapse of O/Ne/Mg white dwarfs while later low mass transients will be similar to stripped-envelope core-collapse supernova}, and may be the last interesting astrophysical transients to occur prior to heat death.  
\end{abstract}




\section{Introduction}

While it is now expected that all stars will evolve toward degenerate remnants (such as neutron stars (NS) or white dwarfs (WD)) or black holes, the fate of these objects in the far future is an open question. Observations of the accelerating expansion and $\Lambda$CDM now suggest that our universe will expand forever, becoming increasingly dark energy dominated while the temperature and matter density asymptotically approach zero. In such a future it is expected that the universe will exhaust all gas for star formation and that almost all stars will become degenerate WDs when they exhaust their fuel over the next $10^{14}$ years \citep{Dyson79,AdamsLaughlin97}.

The ultimate fate of these $\sim 10^{23}$ WDs is an open question.\footnote{Assuming order $10^{12}$ galaxies and $10^{11} $ stars per galaxy in the observable universe.} \citet{Dyson79} and \citet{AdamsLaughlin97} both describe a far future in which all galaxies evaporate through gravitational scattering which ejects most objects, while occasional collisions or encounters with black holes destroy a minority of remnants. Those in binaries merge due to gravitational wave radiation if close, but are more likely disrupted by encounters during the period of scattering that evaporates galaxies. Therefore, virtually all surviving degenerate remnants are eventually isolated on sufficiently long timescales \new{(approximately $10^{20}$ years)}. In the far future it therefore seems likely that a large amount of baryonic matter will be found in degenerate remnants.

Low mass stars, comprising the bulk of all stars today, will be abundant among these remnants having evolved to WDs. \citet{vanhorn68} argues that as WDs cool their cores freeze solid and release latent heat which has recently been observationally confirmed with Gaia \citep{Tremblay}. On cosmological timescales WDs without a heat source will fully crystallize and cool to equilibrium with the cosmic background radiation, at sub-Kelvin temperatures in the far future, though there may be a period of heating due to halo dark matter annihilation of $10^{25}$ up to $10^{47}$ years depending on the exact dark matter candidate considered \citep{AdamsLaughlin97}. Any future evolution of these near absolute zero `black dwarfs' then depends on the stability of the proton. 

If the proton decays then \citet{AdamsLaughlin97} expect these black dwarfs to decay on timescales of $10^{32}$ to $10^{49}$ years. In the alternative scenario, where the proton is stable, we can expect black dwarfs to become iron rich in the far future. Pycnonuclear fusion reactions, driven by quantum tunneling of adjacent nuclei on the crystal lattice within black dwarfs, will tend to process the matter toward iron-56 (which is possibly the ground state of baryonic matter \citep{Page1992}). \citet{Dyson79} estimates a tunneling timescale of

\begin{equation}\label{eq:dys}
T = e^S T_0
\end{equation}

\noindent where $T_0$ is a characteristic nuclear timescale ($\hbar / m_p c^2 \approx 10^{-25}$ s) 
and $S$ is an action integral approximated for fusion by

\begin{equation}
S \approx 30 A^{1/2} Z^{5/6}    
\end{equation}

\noindent with $A$ and $Z$ the mass and charge of the fusion product. \citet{Dyson79} obtains these expressions by approximating the action $S \approx (8 M U d^2 / \hbar^2)^{1/2}$ using mean barrier height $U$ and width $d$ for a tunneling particle of mass $M$. This barrier is then taken to be the Coulomb barrier screened over a distance $d=Z^{-1/3}(\hbar / me^2)$ with a reduced mass $M =  A m_p/4$ for two $A/2$ and $Z/2$ nuclei. This treatment excludes the sensitive density dependence for pycnonuclear fusion, which occurs most quickly at higher densities \cite{SchrammKoonin90,Afanasjev2012,meisel2018nuclear}. For silicon-28 nuclei fusing to iron group elements, Dyson's scheme gives an approximate $10^{1500}$ year timescale for tunneling to process black dwarfs to iron-56.

WDs (and black dwarfs) are electron degenerate which supports matter up to a finite mass limit, the Chandrasekhar mass, given by \citep{chandrasekhar1931highly,chandrasekhar1935highly,maoz2016astrophysics}:

\begin{equation}\label{eq:mch}
    M_{Ch} \approx 1.44 (2 Y_e)^2 M_\odot.
\end{equation}

\noindent For a canonical WD ($Y_e = 0.5$) we obtain the known Chandrasekhar mass limit. If a WD exceeds the Chandrasekhar mass it will collapse \citep{howell2011type,maoz2016astrophysics}. 

As a white dwarf evolves toward a black dwarf, and eventually an iron black dwarf, its equation of state is always that of a relativistic electron gas. However, in the far future when light nuclides are converted to iron-56 the electron fraction of the core will have decreased to $Y_e = 26/56 = 0.464$. Thus, by Eq. \ref{eq:mch} the maximum mass will be smaller by a factor $(0.464/0.5)^2 \approx 0.86$. Therefore, the effective upper mass limit of these degenerates remnants is decreasing with time down toward about $1.2 M_\odot$ as they become increasingly iron rich. Indeed, it is long known in the supernova literature that iron cores above 1.12 $M_\odot$ will collapse \cite{Baron90}.
As black dwarfs above this mass limit accumulate iron in their cores they will necessarily exceed their Chandrasekhar mass precisely because the Chandrasekhar mass is decreasing with $Y_e$. As a result, one may expect catastrophic collapse of these black dwarfs in the far future, powering transients similar to supernova today. 

In this work we consider this fate for the most massive isolated white dwarfs, between approximately 1.2 and 1.4 solar masses. \new{These white dwarfs are the evolutionary endpoint of stars just below the zero-age main sequence mass (ZAMS) for core collapse, roughly 7 to 10 solar masses, and are now thought to be composed of O/Ne/Mg \cite{Nomoto84,Nomoto87,RevModPhys.74.1015,heger2003massive}. There are two phases of evolution to be considered. First, there is a cosmologically long phase of pycnonuclear burning which occurs at near zero temperature which is the primary focus of this work. Once the Chandrasekhar mass is reached, gravitational contraction may proceed on hydrodynamical and nuclear burning timescales, and the transient begins which we only briefly speculate on here.}

\new{
In Secs. \ref{sec:2}, \ref{sec:3}, and \ref{sec:4} we consider the internal structure and determine the mass-radius relation for progenitors at the time of collapse. We calculate the abundance of progenitors in Sec. \ref{sec:5} and their lifetimes in Sec. \ref{sec:6}, and speculate on some general properties of the astrophysical transients associated with collapse in \ref{sec:7}. In Sec. \ref{sec:8} we also consider how the rate and cosmological state of the universe at the time. We conclude in Sec. \ref{sec:9}
}

\section{The Maximum mass of Electron Degenerate Stars}\label{sec:2}

The Chandrasekhar mass can be calculated easily from the ultra-relativistic equation of state, $P \propto \rho^{4/3}$. 
Integrating the structure profile of a WD (\ie\ its density radially outward from the core) yields its mass \citep{chandrasekhar1931highly,chandrasekhar1935highly,maoz2016astrophysics,koonin2018}. For this equation of state one finds that there is a fixed mass for all WDs, regardless of the core density (\ie\ the boundary condition of integration). This equation of state will yield arbitrarily small radii for ever increasing core densities, and the convergence to zero radius at infinite central density is clearly shown by \citet{chandrasekhar1935highly}. This of course is not physical and other physics must set an upper limit on the central density.

As a practical matter there is an upper limit on possible core pressures and densities which is set by the condition for electron capture onto iron-56 nuclei \citep{cardall2008,warren2019constraining}. This capture produces manganese-56 ($Z=25$) and a neutrino and due to odd-even staggering this capture is almost immediately followed by another electron capture to chromium-56 ($Z=24$) whenever it occurs. So while the ultrarelativistic equation of state gives an upper limit on the mass of a degenerate electron gas of fixed $Y_e$, reactions exist in the cores of degenerate remnants which, if the threshold is met, further reduce $Y_e$ thereby triggering collapse. These reactions give us a maximum realizable mass of a cold and isolated degenerate remnant which is slightly below the Chandrasekhar mass. This condition is one of several possible triggers for a supernova progenitor today, though given the low temperatures considered here it is our only relevant mechanism \citep{bethe1979equation,Nomoto87,langanke2014role,warren2019constraining}.

The threshold electron Fermi energy needed to spontaneously drive electron capture can be determined by following \citet{Bahcall64}:

\begin{equation}
    m(Z,A) c^2 + \epsilon_F^{(e)} = m(Z-1,A) c^2 + m_e c^2
\end{equation}

\noindent where $m(Z,A)$ is the mass of the nuclide with $(Z,A)$, $\epsilon_F^{(e)}$ is the electron Fermi energy, and $m_e$ is the electron mass. This is a typical criteria for finding capture layers in accreted neutron star crusts \citep{HZ90a,HZ90b,Fantina2018}. For iron-56 and manganese-56 we find $\epsilon_F^{(e)}=4.207$ MeV. This Fermi energy corresponds to pressures of $P = 5.57 \times 10^{26} \textrm{ dyn/cm}^2$, or equivalently a central mass density of  $1.19\times 10^9 \textrm{ g/cm}^3$. Thus, iron black dwarfs with core densities exceeding this will collapse.

\section{Evolution of Black Dwarfs}\label{sec:3}

We now consider the evolution of a WD toward an iron black dwarf and the circumstances that result in collapse. Going beyond the simple order of magnitude estimates of \citet{Dyson79}, we know pycnonuclear fusion rates are strongly dependent on density so they are greatest in the core of the black dwarf and slowest at the surface. Therefore, the internal structure of a black dwarf evolving toward collapse can be thought of as an astronomically slowly moving `burning' front growing outward from the core toward the surface. \new{This burning front grows outward much more slowly than any hydrodynamical or nuclear timescale, and the star remains at approximately zero temperature for this phase.} Furthermore, in contrast to traditional thermonuclear stellar burning, the later reactions with higher $Z$ parents take significantly longer \new{due to the larger tunneling barriers for fusion}. 

The dwarf will undergo progressive stages of burning likely converting the star from \new{O/Ne/Mg} to heavier symmetric nuclei (\ie\ $Z=N$), likely including silicon-28. Complex fusion pathways may exist depending on the relative timescales for other reactions, which future authors may seek to explore \new{with stellar evolution codes such as MESA using a pycnonuclear reaction network}. For example, $\alpha$-particle exchange (or similar transfer reactions) between adjacent nuclides may be one mechanism for achieving a more uniform lower energy composition when evolving a mixture of light nuclides \citep{Chugunov18}. The lattice structure of the mixture may also impact pycnonuclear fusion rates \citep{caplan2020structure}. Whatever the exact fusion pathway, we do not anticipate any disruption or contraction (or any increase in core pressure) of the white dwarf during this process as $Y_e$ is unchanged. 

Finally, once the core has largely burnt to silicon-28 (or a similar nuclide), we expect it will fuse to produce nickel-56 (or a similar iron group element) which can then decay through positron emissions to iron-56, with $Y_e=0.464$. The annihilation of two electrons from these reactions slowly deleptonizes the star and reduces the core $Y_e$, resulting in contraction which increases core pressure. 

Unlike typical supernova progenitors, this contraction will not heat the black dwarf considerably due to the astronomically long fusion timescales. Furthermore, the reduction in radius is of order ten percent ($Y_e^{\textrm{Fe}}/Y_e^{\textrm{Si}}=0.928$) over timescales we estimated above to be of order $10^{1500}$ years. Any cooling mechanism, such as black-body emission, will proceed much more quickly.
To determine when exactly these objects collapse we must calculate how long it takes to evolve to the critical core pressure for electron capture, which requires us to first know their internal structure. Therefore, we determine the stellar profiles of the progenitors just prior to collapse before proceeding.

\section{Internal Structure of Black Dwarfs}\label{sec:4}

We calculate the profiles of the progenitors from the equation of state for relativistic electrons. This smoothly connects the low density nonrelativistic equation of state $P\propto \rho^{5/3}$ to the ultrarelavistic limit where $P \propto \rho^{4/3}$. We follow the procedure of \cite{koonin2018}. Our equation of state is defined piecewise with $P_{trans}$ being a sharp interfacial transition with a density discontinuity. For pressures greater than $P_{trans}$ we use $Y_e = 0.464$ (for iron-56) and at pressures below we use $Y_e = 0.50$. We are not actually sensitive to whether the outer layers are a pure composition or mixture, or even if there are onion layers as in core collapse supernova (CCSN) progenitors, as an electron gas for $Z=N$ nuclei results in an identical stellar profile; as argued above, stable $Z=N$ nuclei are found up to silicon-28, which fuses to iron group nuclides. 

Integrating the classical equations of stellar structure require the core density as a boundary condition \citep{chandrasekhar1935highly,maoz2016astrophysics,koonin2018}; it suffices to observe that the core density we are interested in is always the critical density for electron capture onto iron-56. 
We therefore have only one free parameter, which is $P_{trans}$. 
Interior to this interface the structure of the star is identical just prior to collapse for all progenitors, having an iron-56 core with a central density of $1.19\times 10^9 \textrm{ g/cm}^3$. Furthermore, this means there is a one-to-one mapping from the mass of the progenitor to the density of the silicon-28 at the interface where we find our burning to front, which will be useful for calculating lifetimes below. As the equation of state stiffens above the transition interface, those with transitions nearest to the core will be the most massive.

\begin{figure}
\centering
\includegraphics[trim=0 0 0 0,clip,width=0.475\textwidth]{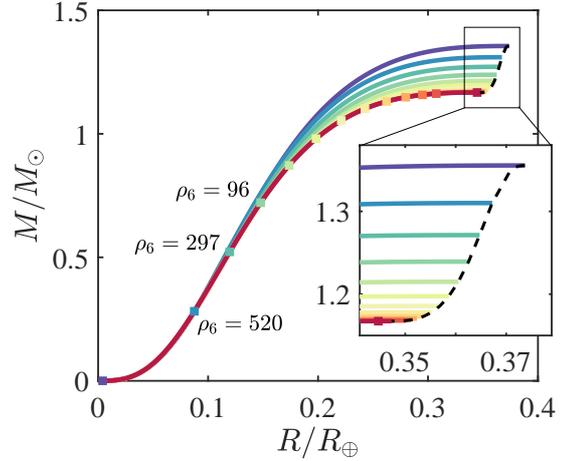}
\caption{\label{fig:wdmr} (Color online)  Profiles of black dwarf supernova progenitors. All profiles have the same central density, and thus the $Y_e=0.464$ part of the profile is identical, following the bottom curve for the pure iron black dwarf (red). The profile is uniquely determined up to some transition pressure at the interface between $Y_e=0.464$ and $Y_e=0.5$ matter (squares). A density discontinuity is found there; we report the density $\rho_{6}$ on the $Y_e =0.5$ side of the interface for our most massive progenitors. At radii above the transition we see the profiles rise off that of a pure iron black dwarf as the equation of state stiffens at higher $Y_e$. In the inset we zoom in on the ends of our curves to show the progenitors mass-radius relationship.}	
\end{figure}

In Fig. \ref{fig:wdmr} we show the mass-radius profiles of progenitor black dwarfs. Each line shows the profile of a progenitor of a different mass just prior to collapse. As the profile of the star within $P_{trans}$ is identical, we plot points (squares) at $P_{trans}$ for each profile to show where it begins to rise off of the profile of a fully iron-56 progenitor (red). For reference, we include the mass density on the silicon side of the interface for the three greatest transition pressures shown, with $\rho_6$ the density in units of $10^6 \textrm{ g/cm}^3$. As expected, the most massive progenitors are approximately $1.35 M_\odot$ and $P_{trans}$ is very near the core while the least massive progenitors are nearly entirely iron-56 and have masses of $1.16 M_\odot$. 

In the inset we show the progenitor mass-radius curve (dotted). Note that the mass of the progenitor is most sensitive to small changes in $P_{trans}$ when it is found at high pressure, which can be seen from profiles with labeled core densities (blue-green). 
For the purely iron black dwarf (red) we find $M=1.167 M_\odot$ and $R=0.347R_\oplus$ while the progenitor with an infinitesimal iron core has $M=1.355 M_\odot$ and $R=0.374 R_\oplus$ (purple). These correspond to the minimum and maximum progenitor masses and radii respectively. As expected both the masses and radii of progenitors only vary by approximately ten percent, of order the difference in $Y_e$ between iron-56 and silicon-28 ($Y_e^{Fe}/Y_e^{Si} = 0.464/0.5 = 0.928$). It can be shown for a relativistic equation of state that $R\propto Y_e$ and $M \propto Y_e^2$, which is what we find here \citep{koonin2018}. 

Our estimate for a fully $Y_e =0.5$ profile is consistent with observations. The most massive WD presently known is REJ 0317-853 whose mass and radius are reported to be $1.35 M_\odot$ and $0.38 R_\oplus$ by \cite{barstow1995re,Burleigh98}; \cite{kulebi2010constraints} gives a mass range of  $1.32-1.38 M_\odot$ due to uncertainty in the core composition, which is consistent with what we find here. 

\new{As a black dwarf is processed toward iron its mass will decrease by releasing energy and neutrinos, though this effect is small and does not affect our treatment above. The burning of alpha cluster nuclei such as O, Ne, and Mg (binding energies $\sim 8$ MeV/nucleon) to iron group elements (binding energies $\sim 9$ MeV/nucleon) during the pycnonuclear phase only releases $\sim 1$ MeV/nucleon. Mass loss during the pycnonuclear burning phase will therefore only reduce the mass of the star by, at most, 1 MeV/939 MeV $\sim 10^{-3}$ which is small. Thus, the mass of the WD is basically equivalent to the mass of the black dwarf supernova progenitor.}

\section{Number of progenitors in the far future}\label{sec:5}

The initial-final mass relation (IFMR) for WDs and a Salpeter initial mass function (IMF) are sufficient to estimate the number of presently observable stars which will evolve to progenitors that may collapse and explode in the far future.

While the initial-final mass relation (IFMR) is now well constrained in the range of $2M_\odot < M_\textrm{initial} <4 M_\odot$, the behavior for $M_\textrm{initial} > 4 M_\odot$ is less well understood. Extrapolating the IFMR from the $2M_\odot < M_\textrm{initial} < 4 M_\odot$ range would tend to underestimate the minimum $M_\textrm{initial}$ as the onset of dredge-up above $4 M_\odot$ tends to reduce the core mass \citep{andrews2015constraints}. For this work, we use the IFMR reported by \cite{cummings2016two} fit to $M_\textrm{initial}>4M_\odot$:

\begin{equation}
    M_\textrm{final} \approx 0.1 M_\textrm{initial} + 0.5 M_\odot. 
\end{equation}

\noindent The behavior of this IFMR is typical for the $M_\textrm{initial}>4 M_\odot$ range and, when extrapolated to higher masses, is the intermediate case when compared to other IFMRs in the literature \citep{cummings2016two,salaris2009semi,catalan2008initial}. This IFMR suggests that stars with a ZAMS mass of $6.5 M_\odot$ or greater will have WD masses greater than the progenitor minimum of $1.16M_\odot$ found above. This is also consistent with measurements of the IFMR with Gaia from \cite{el2018empirical}. Meanwhile, the maximum $M_{\textrm{initial}}\sim 10 M_\odot$ is determined by the minimum ZAMS mass for core collapse supernova progenitors \citep{heger2003massive}. \cite{diaz2018progenitor} use observed supernova remnants in M31 and M33 to determine a minimum ZAMS mass for single-star CCSN progenitors of $7.3 M_\odot$, which we will use as a conservative upper limit. 

The stellar mass density is only expected to grow by about 5 percent above current values \citep{sobral2013large}, so the total number of stars in the universe today represent nearly the total number that will ever exist. Integrating the Salpeter IMF ($\alpha = 2.35$) for $6.5 M_\odot$ to $7.7 M_\odot$ (using $10^{23}$ stars in the observable universe) gives us order $10^{21}$ progenitors. Extending the upper bound to $10 M_\odot$ only increases this by a factor of 2. Order of magnitude uncertainties in the number of stars in the universe prevent us from improving upon this constraint.
\new{As above, stars with a ZAMS in this range are expected to evolve toward O/Ne/Mg WDs \citep{heger2003massive}}

\section{Lifetime of black dwarfs against collapse}\label{sec:6}

We can estimate the lifetime of black dwarfs, and thus the delay time for transients associated with collapse, from the timescale for the \new{pycnonuclear} burning front to grow radially outward. The slowest fusion reaction in the evolution will be the one occurring at the lowest density and with the highest $Z$ parent nuclides. Due the strongly exponential density and $Z$ dependence for pycnonuclear fusion, most of the lifetime of the black dwarf will be spent in the stages just prior to this collapse.

Conveniently, this condition is satisfied by the burning front for the progenitors shown above. Therefore, the fusion rate at the silicon burning front when the core reaches the critical pressure for collapse ultimately determines the lifetime of the black dwarf. Because of this, the most massive black dwarfs will collapse first. Only a small amount of matter already at high density has to fuse to iron-56 before the threshold Fermi energy is reached. Less massive black dwarfs take longer to reach this threshold, requiring ever larger iron cores, while black dwarfs at the lower mass limit only collapse once the outermost layers of the star have fused to iron. 

To obtain a coarse estimate of the lifetime of these progenitors we calculate the lifetime of a silicon-28 nucleus against pyconuclear fusion at the burning front. Following \citet{SchrammKoonin90}\footnote{The factor of $A^2$ in their eq. 33 is corrected to $A$ in the erratum.}, the pycnonuclear fusion reaction rate in the Wigner-Seitz approximation for a bcc crystal is given by

\begin{equation}
    R = (1.06 \times 10^{45}) S \rho A Z^4 \lambda^{7/4} 
    e^{(6.754-2.516\lambda^{-1/2})} \textrm{ cm}^{-3}\textrm{ s}^{-1}  
\end{equation}

\noindent where $S$ is the astrophysical S-factor in MeV barns, $\rho$ is the mass density of the silicon ($A=28$, $Z=14$) at the interface, and $\lambda$ is the ratio of the nuclear Bohr radius and lattice spacing $a$ (for a bcc lattice $\gamma=2$) such that 

\begin{equation}
    \lambda = 0.0245 A^{-4/3} Z^{-2} \gamma^{-1/3} \rho_6^{1/3}
\end{equation}

\noindent which for a bcc silicon-28 lattice evaluates to 
\begin{equation}
\lambda = 1.167 \times 10^{-7} \rho_6^{1/3}. 
\end{equation}

\noindent We obtain the astrophysical $S$-factor following the analytical procedure in \cite{Afanasjev2012} assuming $E=0$, finding $S= 3 \times 10^{63}$ MeV barns. The precise value is unimportant as the rate uncertainty is overwhelmingly dominated by the exponential dependence on density.

The lifetime $\tau$  
of a silicon nucleus at the core-interface can be calculated  using reaction rate in a volume which is the Wigner-Sietz cell, $V_{s} = 2 m/\rho = 4.7 \times 10^{-29} \rho_6^{-1} \textrm{ cm}^3$. As our only free parameter is $\rho_6$, we express the lifetime as 

\begin{equation}
\tau \approx (V_{s}R)^{-1} \approx 10^{83} \rho_6^{-7/12} 
e^{7365 \rho_6^{-1/6} }  \textrm{ s}.
\end{equation}

\noindent For the most massive black dwarfs we find that the core begins fusing to iron in approximately $10^{1100}$ years ($M=1.35 M_\odot$, $\rho_6 = 1.19 \times 10^3$). For an intermediate mass black dwarf approximately half the mass is iron at the time of collapse which occurs in approximately $10^{1600}$ years ($M=1.24 M_\odot$, $\rho_6= 10^2$), comparable to the rudimentary estimate from \cite{Dyson79}. For the least massive black dwarfs the lifetime will be set by the pycnonuclear fusion rates at the surface. Here we expect that the electron gas is not degenerate so atomic silicon may be expected with terrestrial densities for which we find lifetimes of $10^{32000}$ ($M=1.16 M_\odot$, $\rho_6 \sim 10^{-6}$), though at such low densities this treatment may not be appropriate. The exponent of these lifetimes has been rounded to the nearest hundred so our choice of units (years) aren't strictly accurate, however the difference between seconds and the current age of the universe is ultimately small for our purposes.

For comparison, a $10^{11} M_\odot$ black hole (of order the upper limit for supermassive black holes from \citealt{King2015}) has a lifetime due to evaporation from Hawking radiation of $10^{100}$ years. If black dwarf collapse produces transients in the far future, they are likely to be among the last to occur prior to the heat death of the universe.

\section{Black Dwarf supernova transients}\label{sec:7}

\new{Without detailed stellar evolution calculations to determine compositions at the time of collapse it is difficult to make precise predictions about the nature of transients. Nevertheless, order of magnitude arguments and analogs with various presently occurring transients may be informative. For example, we expect that the transients associated with black dwarf supernova will evolve with cosmic time, as the delay time is related to the mass.}

\new{The most massive progenitors will be the first to explode, and thus their composition will be largely unprocessed with only a small iron-56 core, possibly similar to accretion induced collapse of O/Ne/Mg stars. It is an open question whether such explosions produce NSs or low mass WDs with iron cores, as these explosions depend sensitively on the hydrodynamic burning propagation and explosive ignition \citep{Isern1991,canal1992quasi,panei2000evolution,schwab2015thermal}. Recent 3D simulations of electron-capture supernova by \cite{jones2016electron} suggest that an iron core white dwarf might be more likely than a neutron star with analogous progenitors, as in \cite{Isern1991}.}

\new{However, the analog with accretion induced collapse studied in the literature is imperfect. For example, 
given the sensitive dependence of pycnonuclear fusion rates on density and $Z$ it is possible that much of the core has been processed to silicon while a fraction of the lightest nuclei in the outer layers have been processed to heavier nuclei. Therefore, there is less energy available for nuclear burning, especially in the core, and so the transition to a simmering phase, deflagration, and eventual supernova may be much different than currently modeled accretion induced collapse. }

The low mass progenitors will meanwhile have compact iron cores but thin envelopes which may result in dynamics comparable to stripped-envelope core-collapse supernovae (CCSNe) today, and with little accretion after core bounce the shock may expand almost unimpeded. Exact luminosities are difficult to determine without simulations, but the low masses may similarly reduce electromagnetic yields. Compared to CCSNe today we might expect a similar neutrino luminosity as the iron core collapse should be comparable. Gravitational wave emission of the proto-NS may be similar, but without excitation from convection in the infalling matter one might expect less driving of the oscillation. Furthermore, the progenitor is likely highly spherical which will weaken the gravitational wave signal. \new{For all masses considered it seems unlikely that a black hole should form.}  

\section{Transient Rate and Far Future Cosmology}\label{sec:8}

\new{In principle, a volumetric supernova rate can be obtained knowing the number of progenitors as well as their lifetimes from Secs. \ref{sec:5} and \ref{sec:6}. However, given the enormous delay times the consequences of the accelerating expansion of the universe makes it difficult to report this value meaningfully. 

If black dwarf supernova will not begin for another $10^{1100}$ years then it is relevant to consider how much the observable universe will grow in that time from $\Lambda$CDM. As the universe is now entering a dark energy-dominated expansion phase, distances between gravitationally unbound objects will approximately grow like $R(t) \propto e^{H t}$ as in a de Sitter universe (with constant Hubble parameter proportional to some cosmological constant $H \propto \sqrt{\Lambda}$, \citealt{maoz2016astrophysics}). 
Now, recall that \cite{AdamsLaughlin97,Dyson79} show that all orbits become unbound from galaxies (or decay by gravitational wave emission) in only $10^{20}$ years, so we expect all black dwarfs to become isolated with exponentially growing separations relatively soon. Therefore, when black dwarf supernova begin the proper-distance radius of the observable universe will have very roughly grown by $\sim e^{10^{1100}}$ ($H \sim 10^{-11}$ years$^{-1}$ which is in the error of the timescale, discussed above). With such large separations, it may not be intuitive or meaningful to attempt to report a volumetric rate.

Instead, to get a sense of the cosmological conditions where black dwarf supernova are found, consider that the cosmic event horizon for each comoving observer in a de Sitter universe is found at $r_{eh} = c /H \approx 10^{10}$ lightyears \citep{melia2007cosmic,neat2019intuitive}, so after becoming gravitationally unbound we may expect all objects to recede beyond their mutual cosmic event horizons on a timescale similar to the current age of the universe \citep{AdamsLaughlin97}. We therefore expect that every degenerate remnant in the universe will be causally disconnected from every other degenerate remnant, which will all see each other red-shifted to infinity long before any transients considered in this work may occur \cite{maoz2016astrophysics,davis2004expanding}.  
}

\section{Conclusions}\label{sec:9}

Black dwarf supernova progenitors with masses between about 1.16 and 1.35 $M_\odot$ are at near zero temperature which allows us to determine their internal structure with fair accuracy using only a relativistic Fermi gas. Furthermore we can calculate their lifetimes with simple analytical formulas for the pycnonuclear reaction rates at the surfaces of their iron cores. If proton decay does not occur then in the far future we expect approximately one percent of all stars today, about $10^{21}$ stars, to collapse and explode in supernova beginning in approximately $10^{1100}$ years and lasting no more than about $10^{32000}$ years. At such advanced time it is difficult to imagine any other astrophysical processes occurring, which may make black dwarf supernova the last transients to occur in our universe prior to heat death.

Ultimately black dwarfs will explode because pycnonuclear fusion reactions slowly process their interiors to iron, decreasing $Y_e$ and their Chandrasekhar mass. This is amusingly like the inverse of \new{accretion induced supernova today; whereas accretion induced collapse involves a WD increasing its mass up to near the Chandrasekhar limit,} a black dwarf supernova involves a dwarf decreasing its Chandrasekhar limit down to its mass. Unfortunately, due to the timescales involved there are few observable consequences for this work, but this result may nevertheless be of popular interest. However, given the simplicity of the progenitors described in this work it may be interesting and straightforward to \new{simulate the evolution of the structure and composition of progenitors with nuclear reaction networks accounting for pycnonuclear fusion, such as with MESA. Furthermore, with detailed calculations of the evolved compositions from MESA it may be possible to explore these transients in more detail with supernova codes.}

\section*{Acknowledgements}

M.C. thanks Neil Christensen, Zach Meisel, MacKenzie Warren, and the students in the Spring 2020 PHY 308 \textit{Astrophysics} course at Illinois State University for conversation and inspiration. 
This work benefited from support by the National Science Foundation under Grant No. PHY-1430152 (JINA Center for the Evolution of the Elements).

\section*{Data Availability}
The data underlying this article are available in the article.




\bibliographystyle{mnras}








\label{lastpage}
\end{document}